\documentclass[aps,prc,twocolumn,groupedaddress,showpacs,floatfix]{revtex4}

\usepackage{graphics}
\begin{document}


\title{Probing the order-to-chaos region \\
in superdeformed $^{151}$Tb and $^{196}$Pb nuclei with continuum $\gamma$-transitions}



\author{ S.Leoni$^{1}$, G.Benzoni$^{1}$, N.Blasi$^{1}$, A.Bracco$^{1}$, S.Brambilla$^{1}$, F. Camera$^{1}$,
 A.Corsi$^{1}$, F.C.L.Crespi$^{1}$, P. Mason$^{1,2}$, B.Million$^{1}$, D.Montanari$^{1}$, M.Pignanelli$^{1}$,
 E.Vigezzi$^{1}$, O.Wieland$^{1}$, M. Matsuo$^{3}$, Y. R. Shimizu$^{4}$, P.Bednarczyk$^{5}$,
 M. Castoldi$^{6}$, D.Curien$^{7}$, G. G.Duch{\^e}ne$^{7}$, B. Herskind$^8$, M. Kmiecik$^{5}$, A. Maj$^{5}$,
 W. Meczynski$^{5}$, J.Robin$^{7}$, J. Styczen$^{5}$, M. Zieblinski$^{5}$, K. Zuber$^{5}$, A.Zucchiatti$^{6}$ }


\affiliation{$^1$ Dipartimento di Fisica and INFN, Sezione di Milano, Milano, Italy}

\affiliation{$^2$ Laboratori Nazionali di Legnaro, Padova, Italy}

\affiliation{$^3$ Graduate School of Science and Technology, Niigata University, Japan}

\affiliation{$^4$ Department of Physics, Kyushu University, Fukuoka, Japan}

\affiliation{$^5$ The Niewodniczanski Institute of Nuclear Physics, Polish Academy of Sciences, Krakow, Poland}

\affiliation{$^6$ INFN sezione di Genova, Genova, Italy}

\affiliation{$^7$ Institut de Recherches Subatomiques, Strasbourg, France}

\affiliation{$^8$ The Niels Bohr Institute, Copenhagen, Denmark}


\date{\today}

\begin{abstract}
The $\gamma-$decay associated with the warm rotation
of the superdeformed (SD) nuclei $^{151}$Tb and $^{196}$Pb has been measured with the EUROBALL IV array. Several
independent quantities provide a stringent test of the population and decay dynamics in the SD well. A Monte Carlo
simulation of the $\gamma$-decay based on microscopic calculations gives remarkable agreement with the data only
assuming a large enhancement of the B(E1) strength at low excitation energy, which may be related to the evidence
for octupole vibrations in both mass regions.
\end{abstract}


\maketitle

The thermally excited, rapidly rotating atomic nucleus is an ideal laboratory for studying structure properties
beyond mean field. In particular, the $\gamma$-decay which cools the hot compound nucleus at high angular momentum
is a powerful tool for investigating the transition from the chaotic regime around the particle binding energy to
the cold, ordered system close to the yrast line \cite{Bra02,Dos96,Kho98}. This topic, mostly investigated in
normally deformed (ND) nuclei \cite{Bra02,Dos96,Ben05,Leo04,Leo05,Ste05}, becomes particularly interesting and
challenging, both experimentally and theoretically, in the case of very elongated systems. The study of
superdeformed (SD) nuclei requires, in fact, high selectivity and high statistics experiments, in order to focus
on the small fraction  of the $\gamma$-decay ($\approx$ few $\%$) which finally gets trapped into the SD well and
populates discrete rotational bands. At present, there is only partial experimental information available for few
SD nuclei \cite{Leo97,Leo01,Lau07,Ben07,Lop08}, whose interpretation is based on models which are either rather
schematic \cite{Leo01,Ben07,Lop08} or depend on several parameters \cite{Leo97,Lau07}.

The aim of this letter is to make a substantial step forward in the understanding of the population and decay of
warm SD nuclei in the mass regions A=150 and A=190, where superdeformation is a well established phenomenon, by a
comparative and comprehensive study of the nuclei $^{151}$Tb and $^{196}$Pb. For the first time, several
independent observables are extracted from quasi-continuum $\gamma$-coincidence spectra. This provides strong
experimental constraints on the dynamics of the $\gamma$-decay flow and of the tunneling through the potential
barriers between SD and ND excited states, over a wide spin range. A key point of this work is the development of
a new Monte Carlo simulation of the $\gamma$-cascades in the SD well. This makes use of discrete levels calculated
microscopically with the cranked shell model. In addition, the coupling between excited SD and ND configurations
is included, based on penetration probabilities obtained from a microscopical model \cite{Yos01}. It is
anticipated that such a detailed study of the order-to-chaos region reveals the influence of nuclear structure
effects. In particular, an enhancement in the E1 strength around 1-2 MeV, which could be related to octupole
vibrations, is found to play an important role.

\begin{figure}
\resizebox{0.5\textwidth}{!}{\includegraphics{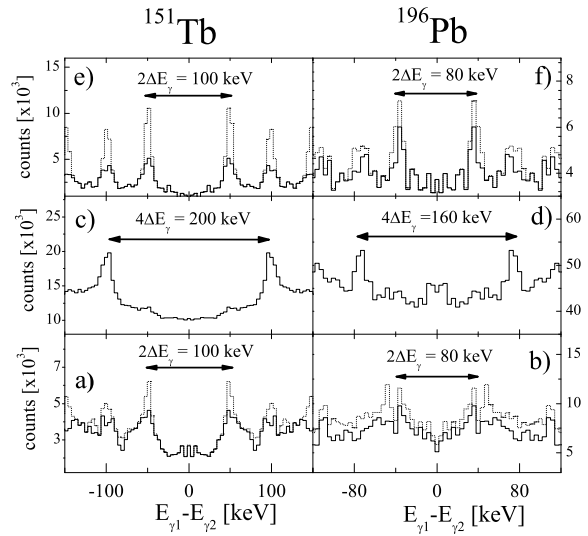}}
\caption{Projections perpendicular to the main diagonal of $\gamma$-coincidence matrices of $^{151}$Tb (left) and
$^{196}$Pb (right), for the energy interval 1150-1280 keV (i.e. $\approx$ 48-56 $\hbar$) and 430-680 keV (i.e.
20-34 $\hbar$), respectively. Panels a), b) show cuts on the Total matrices, c), d) from the rotational planes and
e), f) from the SD-gated spectra (see text for details).} \label{fig1}
\end{figure}

The experiments were performed at the Vivitron in Strasbourg (F) with the EUROBALL IV array \cite{Euroball}. The
reaction used were $^{130}$Te($^{27}$Al,6n)$^{151}$Tb (at 155 MeV) and $^{170}$Er($^{30}$Si,4n)$^{196}$Pb (at 148
MeV). In both cases, a stack of two self-supporting targets (with a total thickness of $\approx$ 1 and 1.2
mg/cm$^2$, respectively) were employed. In the Tb case, the full Ge ball was used, while in the Pb experiment the
low efficiency Ge detectors in the forward hemisphere were replaced by 8 large volume BaF$_2$ scintillators to
measure high energy $\gamma$-rays from the giant dipole resonance (GDR). In both cases an InnerBall of BGO
detectors was used to determine the $\gamma$-multiplicity of every event. The SD nuclei $^{151}$Tb and $^{196}$Pb
were investigated in detail by discrete $\gamma$ spectroscopy \cite{Sin02}. In both cases discrete transitions
linking the SD yrast band to ND configurations were identified, and in the $^{196}$Pb case it was possible to
determine the spin and excitation energy of the SD yrast band \cite{Rob08,Wil05}. In this work we focus instead on
the unresolved excited SD rotational bands, which form ridge and valley structures in $\gamma$-coincidence
spectra. The two data sets have been sorted into a number of $\gamma-\gamma$ matrices in coincidence with
low-lying ND transitions of the isotope of interest ($^{151}$Tb or $^{196}$Pb), named Total, and in coincidence
with the SD yrast band of each nucleus (named SD-gated). In addition, rotational planes (ROT-Plane), namely
matrices obtained selecting triple coincidences with the requirement $x+y=2z$, $x,y$ and $z$ being the energies of
the $\gamma$-rays, were also constructed in coincidence with the nucleus of interest \cite{Leo99}, in order to
enhance the sensitivity to rotational correlation in the $\gamma$-cascades. In all cases a condition on high-fold
events (F $\geq$ 25 for $^{151}$Tb and F $\geq$ 10 for $^{196}$Pb) was imposed, to better focus on
high-multiplicity cascades. For each matrix the background was subtracted by scaling the ungated spectrum by a
factor corresponding to the P/T of the gating transitions and further applying the COR procedure \cite{COR}, in
order to eliminate the remaining uncorrelated background. In all spectra cuts perpendicular to the main diagonal
reveal the existence of ridge structures extending over a broad interval of transition energies (corresponding to
the spin regions 40 to 60 $\hbar$ and 10 to 40 $\hbar$ for $^{151}$Tb and $^{196}$Pb), with a spacing between the
two most inner ridges equal to $2 \Delta E_{\gamma}= 8\hbar^2/\Im^{(2)} \approx$ 100 and 80 keV respectively,
where $\Im^{(2)}$ is the moment of inertia of the SD yrast band in each nucleus. Figure \ref{fig1} shows examples
of such cuts for the wide spin intervals $\approx$ 48-56 $\hbar$ ($^{151}$Tb) and $\approx$ 20-34 $\hbar$
($^{196}$Pb). In the case of the Total and SD-gated spectra (a), b) and e), f)) the dotted lines represent the
full intensity distributions, while the solid lines show the intensity remaining after the subtraction of all
known ND and SD discrete peaks, making use of the RADWARE package \cite{RAD1}. The ROT-planes spectra (c) and d))
are sum of cuts taken in between the SD yrast peaks. In this case, the innermost ridge corresponds to the second
ridge in the Total and SD-gated matrices, and the associated spacing is equal to 4 $\Delta E_{\gamma}$.

\begin{figure}
\resizebox{0.54\textwidth}{!}{\includegraphics{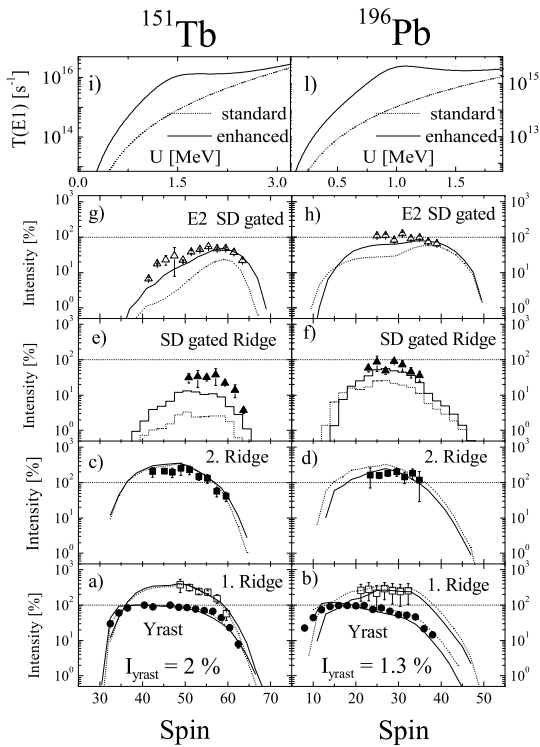}}
\caption{Population of the excited rotational bands in the SD well of $^{151}$Tb (left) and $^{196}$Pb (right),
relative to the corresponding SD yrast (normalized to $100\%$ at the intensity
plateau and shown by circles in the bottom panels). The
symbols show the experimental data: the intensities of the 1$^{st}$ and 2$^{nd}$ ridge are given in panels a), b)
and c), d); panels e), f) refer to the SD-gated ridges, while panels g), h) show the total E2 population in
coincidence with the SD band, as extracted from the E2 bump (see text). Predictions from simulations of the
$\gamma$-decay assuming a standard (enhanced) E1 strength (see top panels i) and l)) are shown by dotted (solid)
lines. } \label{fig2}
\end{figure}

In first place, the intensity of the ridge structures has been evaluated and compared to the SD yrast, as shown by
symbols in figure \ref{fig2}. In both nuclei the total intensity of the 1$^{st}$ and 2$^{nd}$ ridge (panels a), b)
and c), d)) is up to 3 times larger than the SD yrast population at the plateau (which collects 2$\%$ and $1.3\%$
of the total decay-flux, respectively), pointing to the existence of several discrete unresolved SD bands decaying
directly into the low deformation minimum. The more selective analysis of the 1$^{st}$ ridge in coincidence with
the SD yrast (panels e) and f)) shows, in fact, that only a fraction of the total population of the discrete
excited bands is finally collected into the SD well, corresponding at most to 40$\%$ and 80$\%$ of the SD yrast
intensity in $^{151}$Tb and $^{196}$Pb, respectively. Finally, the symbols in panels g) and h) give the total
intensity of the rotational decay in coincidence with the SD yrast, including also the contribution from
fragmented/damped bands at higher excitation energies. This is obtained by the analysis of the E2 bump observed in
one dimensional spectra, in coincidence with the SD band, which in the case of Pb show the presence of
contaminants (probably M1's) in the spin region 20-30 $\hbar$ \cite{Leo08}. It is found that the intensity of the
SD-gated ridges accounts for almost the entire E2 strength up to spin 55 $\hbar$ in $^{151}$Tb and 35 $\hbar$ in
$^{196}$Pb, corresponding to the plateau regions of the SD yrast, while at higher spins the ridge intensity drops
to less than half of the E2 bump, indicating that most of the feeding of the SD yrast comes from fragmented
rotational bands in the region where rotational damping largely dominates \cite{Bra02}. This differs from the
peculiar case of $^{194}$Hg, were an exceptionally narrow ridge, exhausting nearly all E2 decay strength has been
interpreted as a signature of ergodicity in a nuclear system \cite{Lop08}.

\begin{figure}
\resizebox{0.52\textwidth}{!}{\includegraphics{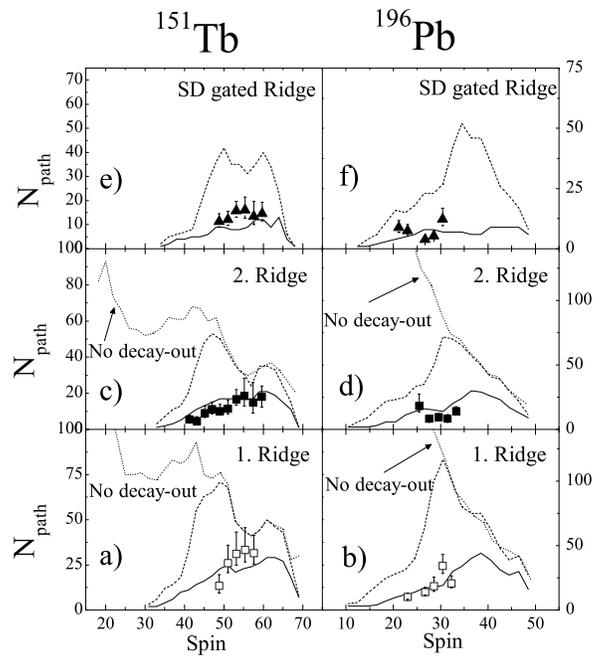}}
\caption{The number of discrete excited SD bands obtained from the analysis of the count fluctuations of the ridge
structures of $^{151}$Tb (left) and $^{196}$Pb (right). Panels a), b) and c),d) refer to the 1$^{st}$ and 2$^{nd}$
ridge, e), f) to the SD-gated ridge. Symbols refer to data, solid lines to the Monte Carlo simulation. See the
text for the dashed and dotted lines.}
\label{fig3a}
\end{figure}

Additional information on the dynamics of the $\gamma$-decay in the SD well and on the coupling with the ND states
can be obtained estimating the number of discrete bands populating the ridge structures from the
analysis of the fluctuations of the counts in the $\gamma$-coincidence matrices \cite{Dos96,Bra02}. The results
are shown by symbols in figure \ref{fig3a} for $^{151}$Tb (left) and $^{196}$Pb (right). In both cases
a rather large number of SD discrete bands (up to more than 30) are found to populate the Total ridges (a) and
b)), half of which extend at least  over 3 consecutive transitions, as follows by the analysis of the second
ridge (c) and d)). In addition, the study of the SD-gated ridges (e) and f)) show that no more than half of the
discrete excited bands are feeding into the SD yrast, supporting the analysis of the ridge intensities, previously
discussed.

The experimental findings have been compared with predictions based on a Monte Carlo simulation of the
$\gamma$-decay flow of $^{151}$Tb and $^{196}$Pb. The code is  an extended version of MONTESTELLA \cite{Bra96} and
simulates the $\gamma$-decay of the excited rotating nucleus from the residual entry distribution towards the SD
or the ND yrast line. It is based on the competition between E2 collective and E1 statistical transitions in both
minima and on a tunneling probability across the barrier separating the two potential wells, accordingly to the
equations of ref. \cite{Sch89}. In the SD well energy levels, E2 transition probabilities and potential barriers
are microscopically calculated according to the cranked shell model of ref. \cite{Yos98,Yos01}, including, in the
case of Pb, the phenomenological adjustment of the mass parameter $C_{mass}= 2.7$ to reproduce the decay-out spin
of the SD yrast. At excitation energies higher than the ones covered by the microscopic states (i.e. U$>$ 4.5 and
3.5 MeV in $^{151}$Tb and $^{196}$Pb, respectively) the rotational decay in the SD well proceeds through a
continuum of states, governed by level densities, E2 strengths (with a width $\Gamma_{rot} \approx$ 100, 70 keV,
respectively) and barriers extrapolated from the region of microscopic levels. The $\gamma$-decay in the ND well
is instead described schematically, as done in the simulations of ref. \cite{Leo97,Lau07}, with a density of
states calculated accordingly to ref. \cite{Gor01}. Once the entry distribution of the $\gamma$-decay is defined,
the use in the SD well of microscopically calculated quantities leaves no room for adjustable parameters, except
for the E1 transition probability, which is described in terms of the Lorentzian strength function of the giant
dipole resonance, with an hindrance factor tuned to reproduce the intensity of the SD yrast (as in the code
MONTESTELLA \cite{Bra96,Sch89}). The residual entry distributions for the $\gamma$-decay of $^{151}$Tb and
$^{196}$Pb have been calculated in two steps. Firstly, we have used the heavy ions-collision model of ref.
\cite{Win95}, which provides the fusion cross section of the compound nuclei $^{157}$Tb and $^{200}$Pb. Secondly,
we have determined the neutron evaporation by the Monte Carlo version of the code CASCADE \cite{Pul77}, obtaining
angular momentum and excitation energy distributions of the $\gamma$-decay with a maximum at I = 60 and 40 $\hbar$
and U = 8.3 and 7.7 MeV above yrast, for $^{151}$Tb and $^{196}$Pb, respectively.

The Monte Carlo simulation provides predictions for all the observables shown in figure \ref{fig2} and
\ref{fig3a}, including, for the first time, the intensity and number of discrete bands in coincidence with the SD
yrast, which depend strongly on the competition between E2 and E1 transitions along the cascades. For both nuclei
the model reproduces rather  well the more inclusive quantities, such as the SD yrast intensity and the population
of the discrete excited bands forming the 1$^{st}$ and 2$^{nd}$ ridge (dotted lines in figures 2 a), b) and c),
d)). On the other hand, it largely underestimates the E2 intensity observed both in the ridge and E2 bump in
coincidence with the SD yrast (dotted lines in panels e),f) and g),h)). These observables, although referring to
the warm rotation in the transition region between order and chaos, are in fact very sensitive to the delicate
balance between E2 and E1 transitions at low excitation energy, where nuclear structure effects play a major role.
It is worth noticing that in both A=150 and A=190 mass regions, experimental evidence has been found for octupole
vibrations in superdeformed nuclei ($^{152}$Dy \cite{Lau02}, $^{196}$Pb \cite{Ros01}, $^{190}$Hg \cite{Kor01}),
resulting in strongly enhanced E1 transitions, linking the excited SD bands to yrast. The B(E1) values are in fact
of the order of 10$^{-4}$ W.u., namely 1 to 2 orders of magnitude larger than expected from the tail of the GDR,
in agreement with various theoretical models \cite{Kva07,Nak95}. We have therefore performed simulation
calculations for $^{151}$Tb and $^{196}$Pb making use of a GDR strength with an extra component giving rise to an
enhancement in the B(E1) values between 10 to 100, in the excitation energy region around 1.5 and 1 MeV,
respectively, i.e. consistent with the energies of the E1 enhanced transitions experimentally observed (see figure
2 i) and l)). The obtained results (solid lines in figure 2) are now much better in agreement with the data,
especially for the E2 strength in coincidence with the SD yrast (panels e), f) and g), h)). The number of paths
$N_{path} =1/{\sum w_i^2}$, where $w_i$ indicates the relative population $w_i$ of each path, is also in good
agreement with the data (solid lines in Fig. \ref{fig3a}). This quantity represents a crucial test for various
aspects of our model. $N_{path}$ gives the number of discrete bands which are populated sufficiently strongly and
it is determined by the level density, by the onset of damping (which takes place at $U \sim 2 $ MeV in $^{196}$Pb
and at $U \sim 3 $ MeV in $^{151}$Tb) and by the population mechanism. The latter favours low-lying bands around
1-1.5 MeV, not only because they tunnel less easily through the potential barrier but also because they receive
stronger E1 feeding. We note that $N_{path}$ is much lower than the number of discrete bands $N_{band}$ obtained
from the cranked shell model ignoring the flow (dotted lines in Fig. \ref{fig3a}): taking into account the
depopulation of the bands due to the tunneling (as estimated in ref. \cite{Yos01}) decreases $N_{band}$  at low
spins but still largely overestimates the data (dashed lines). This is different from the case of ND nuclei, where
discrete bands are more equally populated and $N_{path}$ is closer to $N_{band}$ \cite{Bra96}.

In summary, we have presented a study of the warm rotation in the SD nuclei $^{151}$Tb and $^{196}$Pb, providing a
series of independent quantities which test our understanding of the $\gamma$-decay in the SD well over the whole
spin range, i.e from the feeding region, down to the decay-out into the ND minimum. A newly developed Monte Carlo
code, based on microscopic quantities, is able to reproduce all the observables except the E2 strength gated on
the SD yrast line. Agreement with this quantity can be obtained only increasing the E1 strength at low excitation
by almost two orders of magnitude, as compared to the standard Lorentzian parametrization for SD nuclei. This may
be related to the observation of strong E1 transitions between discrete bands, observed in both regions and
associated with octupole vibrations. Therefore, the study of the warm rotation in the transition region between
order and chaos not only is instrumental in shedding light on nuclear structure properties beyond mean field, but
it can also be considered as a tool to unravel microscopic features of the $\gamma$-decay.

This work was partially supported by the EU (Contract No. EUROVIV: HPRI-CT-1999-00078), by the Polish Ministry of
Science and Higher Education (Grant Nr. 1-P03B-030-30).


\begin{thebibliography}{19}
\bibitem{Bra02}A. Bracco and S. Leoni, Rep. Prog. Phys. 65, 299(2002)
\bibitem{Dos96}T.D{\o}ssing et al., Phys. Rep. 268, 1(1996)
\bibitem{Kho98}T.L. Khoo, in Tunneling in Complex Systems, ed. S. Tomsovic,
World Scientific (1998), p. 229
\bibitem{Ben05}G. Benzoni et al., Phys. Lett. B615, 160(2005)
\bibitem{Leo04}S. Leoni et al., Phys. Rev. Lett. 93, 022501(2004)
\bibitem{Leo05}S. Leoni et al., Phys. Rev. C72, 034307(2005)
\bibitem{Ste05}F.S. Stephens et al., Phys. Rev. Lett. 94, 042501(2005)
\bibitem{Leo97}S. Leoni et al.,Phys. Lett. B409, 71(1997)
\bibitem{Leo01}S. Leoni et al.,Phys. Lett. B498, 137(2001)
\bibitem{Lau07}T. Lauritsen et al., Phys. Rev. C75, 064309(2007)
\bibitem{Ben07}G. Benzoni et al., Phys. Rev. C75, 047301(2007)
\bibitem{Lop08}A. Lopez-Martens, Phys. Rev. lett. 100, 102501(2008)
\bibitem{Yos01}K. Yoshida et al., Nucl. Phys. A696, 85(2001)
\bibitem{Euroball}J. Simpson, Z. Phys. A358,139(1997)
\bibitem{Sin02}B. Singh et al., "Table of Superdeformed Nuclear Bands and Fission Isomers"(2002).
\bibitem{Rob08}J.Robin et al., Phys. Rev. C77, 014308(2008).
\bibitem{Wil05}A. Wilson et al., Phys. Rev. Lett. 95, 182501(2005)
\bibitem{Leo99}S Leoni et al., Eur. Phys. J. A4, 229 (1999)
\bibitem{COR}O. Andersen et al., Phys. Rev. Lett. 43, 687(1979)
\bibitem{RAD1}D.C.Radford, Nucl.Instr.Meth. A361, 297(1995)
\bibitem{Leo08}S. Leoni et al., in preparation
\bibitem{Bra96}A. Bracco et al., Phys. Rev. Lett. 76, 4484(1996)
\bibitem{Sch89} K. Schiffer et. al., Z. Phys. A332, 17(1989)
\bibitem{Yos98}K. Yoshida et al., Nucl. Phys. A636, 169(1998)
\bibitem{Gor01} S. Goriely et al., At. Data Nuc. Data Tab. 77, 311(2001)
\bibitem{Win95}A. Winther, Nucl. Phys. A594, 203(1995)
\bibitem{Pul77}F. Pulhofer, Nucl. Phys. A280, 267 (1977)
\bibitem{Lau02}T. Lauritsen et al., Phys. Rev. Lett. 28, 282501-1(2002)
\bibitem{Ros01}D. Ro{\ss}bach et al., Phys. Lett. B513, 9(2001)
\bibitem{Kor01}A. Korichi et al., Phys. Rev. Lett. 86, 2746(2001)
\bibitem{Kva07}J. Kvasil et al., Phys. Rev. C75, 034306(2007)
\bibitem{Nak95}T. Nakatsukasa et al., Phys. Lett. B343, 19 (1995)
\end{thebibliography}
\end{document}